\documentclass[prl,twocolumn,aps,superscriptaddress,showpacs]{revtex4-1}
\usepackage[latin2]{inputenc}
\usepackage{graphicx}
\usepackage{color}
\usepackage{epsfig}
\usepackage{amsmath}
\usepackage{amsfonts}
\usepackage{amssymb}

\begin{document}

\title{Minimal energy ensemble Monte Carlo for the partition function of fermions coupled to classical fields}

\author{Przemys{\l}aw R. Grzybowski}
\affiliation{Faculty of Physics, Adam Mickiewicz University in Pozna{\'n}, Umultowska 85, 61-614 Pozna{\'n}, Poland}

\author{{\L}ukasz Czekaj}
\affiliation{National Quantum Information Centre of Gda{\'n}sk, 81-824 Sopot, Poland}

\author{Mariusz Nogala}
\affiliation{Faculty of Physics, Adam Mickiewicz University in Pozna{\'n}, Umultowska 85, 61-614 Pozna{\'n}, Poland}

\author{Adam {\'S}cibior}
\affiliation{Faculty of Physics, Adam Mickiewicz University in Pozna{\'n}, Umultowska 85, 61-614 Pozna{\'n}, Poland}

\author{Ravindra W. Chhajlany}
\affiliation{Faculty of Physics, Adam Mickiewicz University in Pozna{\'n}, Umultowska 85, 61-614 Pozna{\'n}, Poland}
\affiliation{ICFO-Institut de Ci\`encies Fot\`oniques, Av. Carl Friedrich
Gauss 3, 08860 Barcelona, Spain}

\begin{abstract}

Models of non-interacting fermions coupled to auxilliary classical degrees of freedom are relevant to the understanding of a wide variety of problems in many body physics,  {\it e.g.} the description of manganites, diluted magnetic semiconductors or strongly interacting electrons on lattices. Monte Carlo sampling 
over the classical fields is a powerful, yet notoriously challenging, method for this class of problems -- it requires the solution of the fermion problem for each classical field configuration. Conventional Monte Carlo methods minimally utilize the information content of these solutions by extracting single temperature properties. We present a  flat-histogram Monte Carlo algorithm that simulates a novel statistical ensemble which allows to acquire the full thermodynamic information, {\it i.e.} the partition function at all temperatures, of sampled classical configurations.

\end{abstract}

\pacs{05.10.Ln, 02.70.Ss}

\maketitle

{\it Introduction - }
Bilinear Hamiltonians of lattice fermions coupled to classical degrees of freedom (continuous or discrete) are ubiquituous in contemporary many-body physics. These often  arise as suitable approximations in the description of systems where many different degrees of freedom contrive to yield complex and interesting physics. In these cases, some subsystem is treated classically as {\it e.g.} localized spins in double-exchange models \cite{Zener,Anderson,deGennes}, models of Mn-doped (III, IV) semiconductors  \cite{Schliemann}, the Ising t-J model \cite{IsingtJ},  adiabatic phonons in polaron models \cite{Holstein,Millis,Freericks1,Freericks2}, or one species of fermions  in the Falicov-Kimball model \cite{Freericks2,Hubbard,FK}. Exactly solvable models can also take this form, such as the seminal honeycomb lattice Kitaev model \cite{Kitaev} where interacting spins  are mapped onto Majorana fermions coupled to static gauge fields. 
More generally, auxiliary field methods, {\it e.g.} based on the Hubbard-Stratonovich transformation, allow to decouple fermion-fermion or fermion-boson interactions -- the fields are then  treated classically in conjunction with the application of the Suzuki-Trotter decomposition \cite{BSS,HirschFye} (see \cite{Mukherjee14a} for a recent approximate scheme with static fields).

The simplicity of the form of such models belies their complexity. Although obtaining eigenstates for fixed configurations of classical fields is computationally easy, summing over the exponential number of such configurations to obtain thermodynamic properties is notoriously expensive. An obvious approach towards this problem is via Monte Carlo (MC) sampling \cite{Ulam}. Summing over the fermion states for fixed classical field configuration yields the conditional (grand) partition function. This quantity, at fixed temperature, serves as the Metropolis weight \cite{Metropolis} for performing a random walk in the space of classical field configurations \cite{Dagotto-review}. The serious bottleneck in these simulations is the repeated performance of the fermionic trace -- and so exact diagonalization ED of the free fermion system -- for executing the walk. Hence various improvements have been proposed to optimize the reevaluation of the weight - moment expansion of the fermion density of states by Chebyshev polynomials \cite{KPM,TKPM}, low-rank matrix updates  \cite{LRMUp} or  Green's functions  \cite{GFKMP} Chebyshev expansion. On the other hand it seems naturally essential to optimize the extraction of information at each MC step. Indeed, while the expensive ED yields  the conditional partition function at all temperatures, these are completely discarded by using only single temperature data in the above approaches. Here, we introduce an algorithm that {\it fully exploits the thermodynamic information} available at each diagonalization step in MC simulations of  free  fermions coupled to classical fields (FCCF).

The paradigmatic Metropolis algorithm \cite{Metropolis}  suffers from critical slowing down at continuous phase transitions and prolonged trapping in metastable states at discontinuous transitions. These can be overcome to a large extent using cluster update schemes
\cite{Swendsen,Wolff,loop,worm,loopoper,directloop} or  sampling extended ensembles \cite{singlehist,multican,broad,paralel} such as  Wang-Landau sampling of the density of states \cite{WL,QWL}.
 While these approaches have been used in the study of classical and quantum systems, FCCF hold their own system-specific challenges rendering such applications difficult.   
In particular, the effective Hamiltonian  (corresponding to the energies in the Metropolis weight) of the classical fields 
in general contains temperature dependent, long-range, multi-particle  interactions. Often, molecular dynamics (MD) or hybrid Monte Carlo methods using Langevin's equation \cite{hybrid1,hybrid2,hybrid3,molecular} are better suited than standard MC simulations of such Hamiltonians although the acceptance rate in these simulations crucially depends on the quality of the approximate action. Instead, we sample \textit{a novel extended ensemble} bringing the advantages of Wang-Landau-like sampling to FCCF.

{\it The problem and  method - } 
 We first generalize our considerations: a system is {\it bi-separable} if it may be separated into two subsystems $A$ and $B$ such that for a given state of $A$ all states of the system $B$ can be efficiently summed over to obtain the conditional (grand) partition function $Z(\beta|A)$ or equivalently the conditional free energy (grand potential) $F(\beta|A)$. This definition covers both FCCF (where subsystem $A$ is classical) and many classical models (as {\it e.g.} the Ising model on a bipartite lattice where $A$ and $B$ are the spins on the two sublattices respectively). For bi-separable systems, the partition function is decomposed as:
\begin{gather}
Z(\beta)= {\rm Tr}\exp(-\beta H) = \sum_A \big(\sum_B \exp(-\beta H_{B|A}) \big) =
\nonumber\\
 \sum_A Z (\beta |A) = \sum_A \exp(-\beta F (\beta|A))
\label{PartitionFunction}
\end{gather}
i.e. the partition function $Z(\beta)$ is obtained by averaging the conditional partition functions over all configurations attainable in the system $A$. Notice that once the conditional energy spectrum is obtained, complete thermodynamic information associated with the exponentially large number of configurations of $B$ encoded in $\exp(-\beta F (\beta|A))$ for all inverse temperatures $\beta$ becomes potentially available at each simulation step. 

Our {\it problem} can be stated as follows. Standard MC sampling for FCCF systems consists of walking, at fixed temperature, between different configurations of $A$ with with Metropolis weights  $\exp(-\beta F (\beta|A))$. Here, we aim to 
obtain the entire partition function (\ref{PartitionFunction}) by acquiring, at a given simulation step, the conditional partition function $Z(\beta|A)$ for {\it all} arguments  $\beta$ from the information (full spectrum) abundantly 
available for each fixed configuration on $A$.
In principle this may also be achieved by parallel tempering (to efficiently sample configurational space of $A$) and a reweighting procedure, which however requires well chosen set of temperature intervals. Instead, in our method, we  perform a random walk directly in the configurational space of $A$ without referring to any specific temperature. The basic challenge here is to obtain an appropriate importance sampling scheme over the (exponential number of) configurations of $A$. 

The key behind any thermodynamic Monte Carlo simulation  lies in sweeping through energy space efficiently. A simple observation reveals that, for most systems, only a few configurations of $A$ will lead to the spectrum of system energies containing the ground state energy. These configurations of $A$ obviously must be effectively found by the importance sampling scheme. On the other hand the conditional DOSes $\rho_A(E)$ of energy spectra associated with typical configurations of $A$ are expected to differ most also in their lower range. Therefore a key discriminant for configurations of $A$ is the {\it minimal energy attainable by system at a given configuration} which we will denote as $\varepsilon_{\min}(A)$. We consider two configurations of $A$ as belonging to the same {\it class} when they have equal values of $\varepsilon_{\min}$.

Formally, any importance sampling scheme can be represented by assigning a weight function $w(A)$ to the set of configurations of subsystem $A$, such that
\begin{gather}
Z(\beta) = \sum_A w(A) \Bigg( \frac{\exp(-\beta F (\beta|A))}{w(A)} \Bigg) .
%\nonumber\\
%= \Big(\sum_A 1\Big) \times \sum_A w(A) \Bigg( \frac{\exp(-\beta F (\beta|A))}{w(A)} \Bigg) \Bigg/ \sum_A w(A) \times 1/w(A)
%\nonumber\\
%= \Big(\sum_A 1\Big)\times  \Bigg\langle   \frac{\exp(-\beta F (\beta|A))}{w(A)} \Bigg\rangle \Bigg/ \Bigg\langle \frac{1}{w(A)} \Bigg\rangle. 
\label{ImportanceSamplingDef}
\end{gather}
The principle that we identify and implement for acquiring the partition function is that {\it all minimal energy classes be visited by the algorithm}. 
For this, notice that the  minimal attainable energies $\varepsilon_{\min}$ can be associated to a density of states -- ${\tilde \rho} (\varepsilon_{\min})$, which enumerates the number of configurations of subsystem $A$ attaining a given minimal energy. We emphasize that this "auxiliary" DOS is distinct from the true DOS of the system, and does not determine the latter.

Our algorithm  consists of two separate sampling stages associated with first generating the weight distribution for importance sampling and then utilizing it to acquire thermodynamic information about the system. (i) The auxiliary DOS ${\tilde \rho} (\varepsilon_{\min})$ is readily obtained by performing a  Wang-Landau (WL) algorithm \cite{WL} in the space of minimal energies, where $\varepsilon_{\min}$ plays the role of "energy" of a given configuration of $A$. 
(ii) Next, one performs a random walk in the space of configurations of $A$ with weight function 
\begin{gather}
w(A)= 1/{\tilde \rho}(\varepsilon_{\min}(A) )
\label{ourweight}
\end{gather} 
to sample $Z(\beta)$ (Eq.(\ref{ImportanceSamplingDef})) for all $\beta$. This choice of weight function not only allows to visit all classes of configurations but visits each class approximately the same number of times yielding a flat histogram.

The above principle can be viewed as a {\it minimal necessary requirement} for sampling energy space. It yields a coarse-grained view of energy space disregarding any differentiation between configurations belonging to a given class. Below, we show that the application of this principle leads to remarkably accurate results.

{\it Testing the algorithm - } 
The convergence properties of the WL algorithm which constitutes the first stage of our algorithm have been widely studied. 
We note, that the random walk in the second stage of our algorithm (with fixed auxiliary DOS) fulfills, {\it by fiat}, detailed balance. 
	We test the convergence of our algorithm on the (a) Ising and (b) Potts model. Finally, we present continuous temperature results for the Falicov-Kimball model as a prominent example of FCCF.

\begin{figure}[t]
   \centering
   \includegraphics[width=0.85\columnwidth]
  {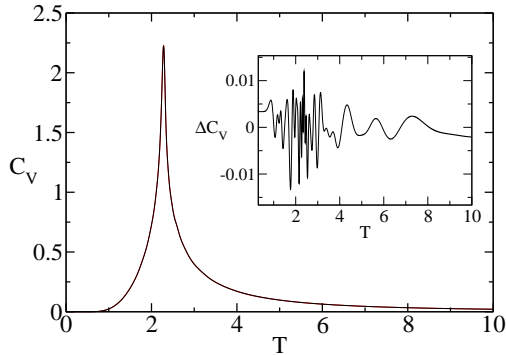}
   \caption{(Color Online). The specific heat $C_V$ per site for the 2D Ising model obtained with our (black line) and WL (red line) algorithms. Lattice size is 60x60 with PBC. Inset shows the relative difference $\Delta C_V = (C_{V1} - C_{V2})/C_{V1}.$  }
    \label{fig1}
 \end{figure}

The Ising model with nearest neighbour interactions on a square lattice  is an ideal benchmark for testing new algorithms since both an exact solution in the thermodynamic limit  as well as large system high precision Monte Carlo results are available. This model is bi-separable which allows for direct application of our algorithm. Indeed for given configuration of spins on one of the sublattices (subsystem $A$) the conditional partition function corresponding to all configurations of spins from the second sublattice (subsystem $B$) is simply:
\begin{gather}\label{ZIsing}
Z (\beta |A) =2^{N/2} (\cosh (2\beta J))^{N_2}(\cosh (4\beta J))^{N_4},
\end{gather}
where $J$ is the coupling between spins, $N/2$ is the number of sites on sublattice $B$. $N_2$ and $N_4$ are the number of spins on sublattice $B$ subject to the nearest neighbour fields with absolute value $2 J$ and $4 J$ respectively. Of course, $N_2$ and $N_4$ are dictated by the configuration of spins on sublattice $A$. Fig. \ref{fig1} shows the continuous temperature dependence of the specific heat, which is the temperature derivative of the free energy, obtained with our and Wang-Landau algorithms. These are generated for a 60 x 60 site lattice (with periodic boundary conditions PBC). The two curves show good agreement, indicating sufficiency of our prescribed importance sampling principle. Any essential problems regarding the effectiveness of this sampling scheme would have been expected to be apparent for this moderately large lattice size. 

\begin{figure}[t]
   \centering
   \includegraphics[width=0.85\columnwidth]
  {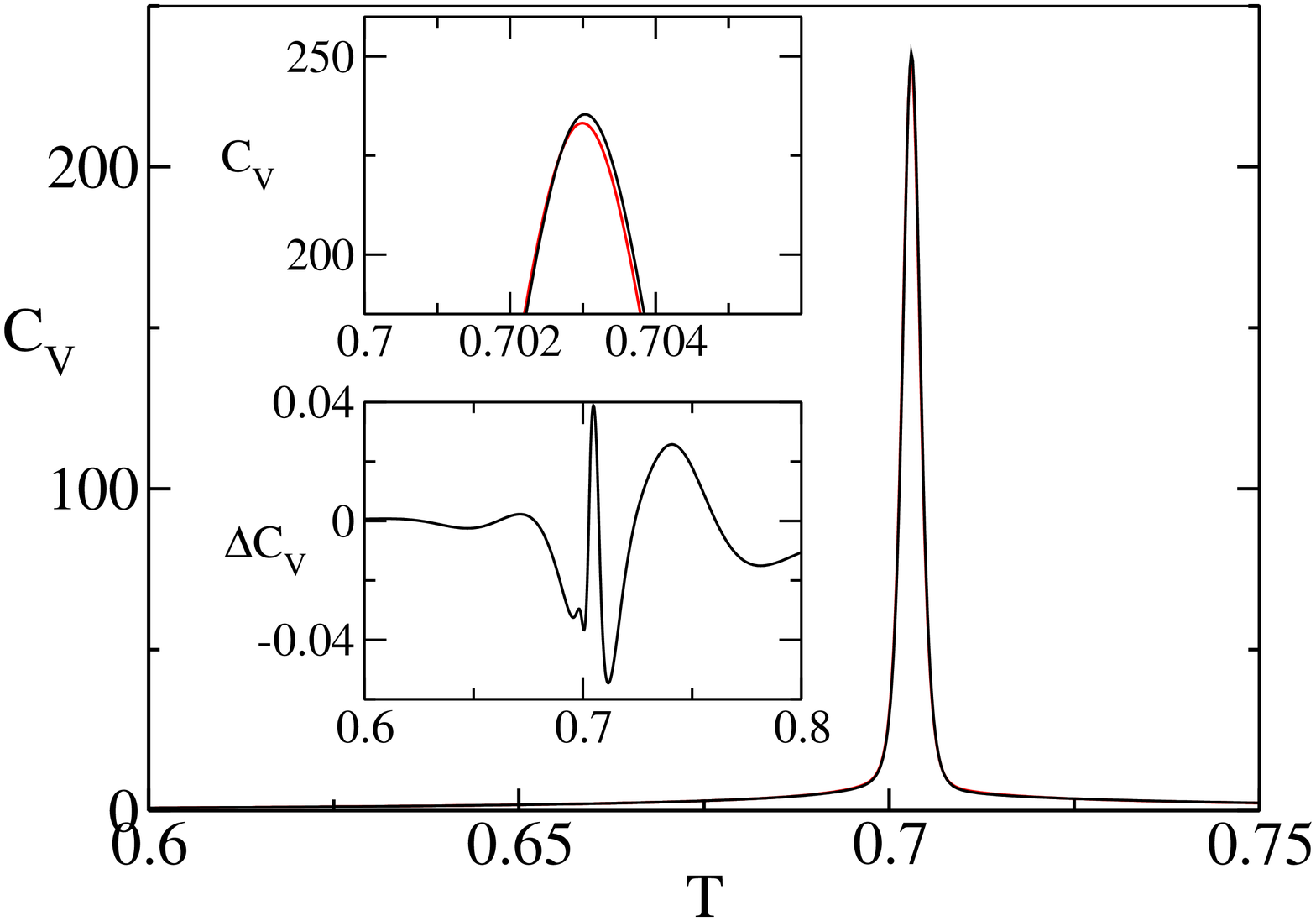}
   \caption{(Color Online). The specific heat $C_V$ per site for the 2D $10$--state Potts model obtained with our (black line) and WL (red line) algorithms. Lattice size is 30x30 with PBC. Upper inset --   tips of the $C_V$ curves from main panel. Lower inset -- relative difference $\Delta C_V = (C_{V1} - C_{V2})/C_{V1}.$ }
    \label{fig2}
 \end{figure}

\begin{figure*}[t]
\centering
\begin{minipage}{.32\textwidth}
  \centering
  \includegraphics[width=1.1\columnwidth]{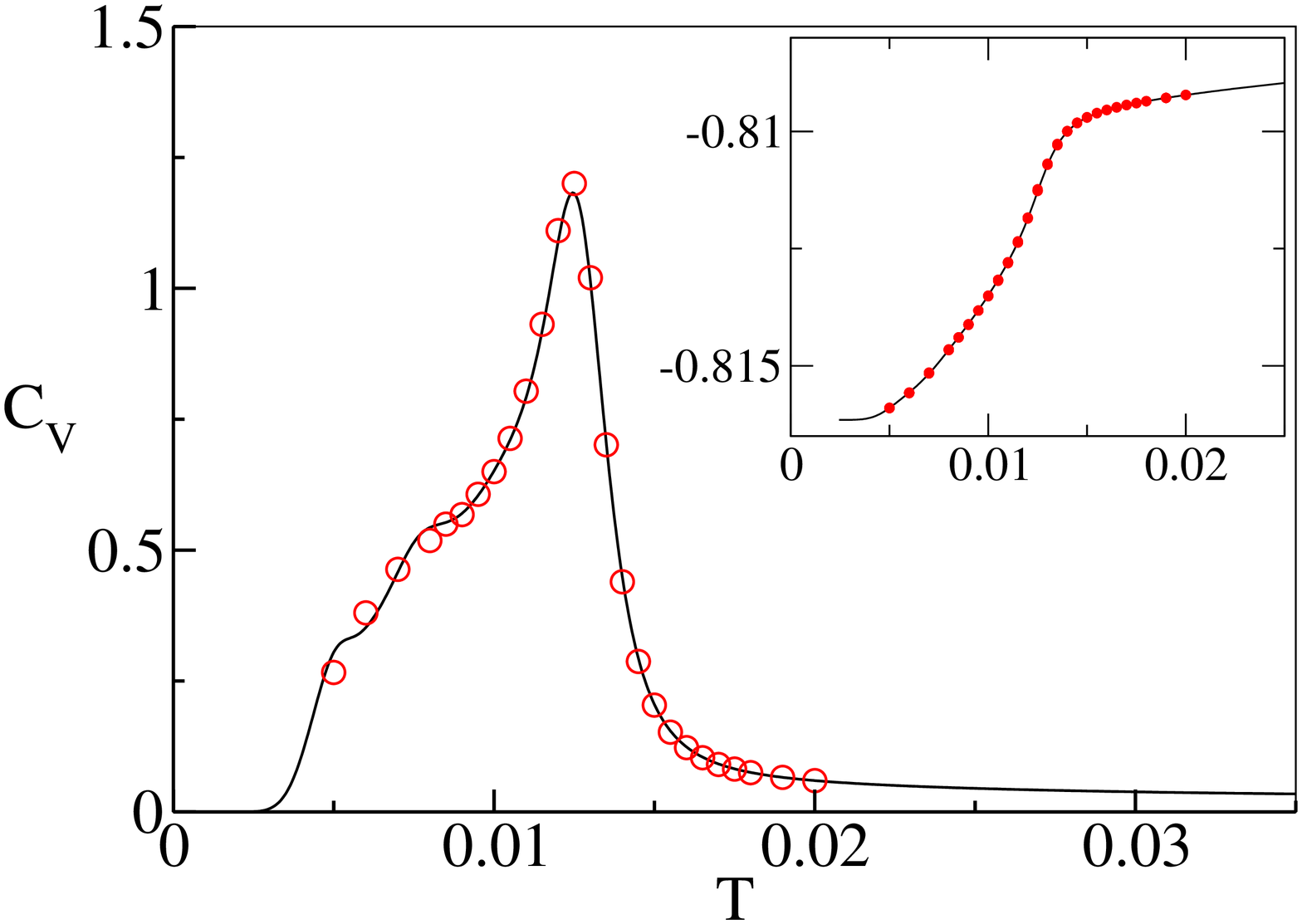}
\end{minipage}%
\begin{minipage}{.32\textwidth}
  \centering
   \includegraphics[width=1.1\columnwidth]{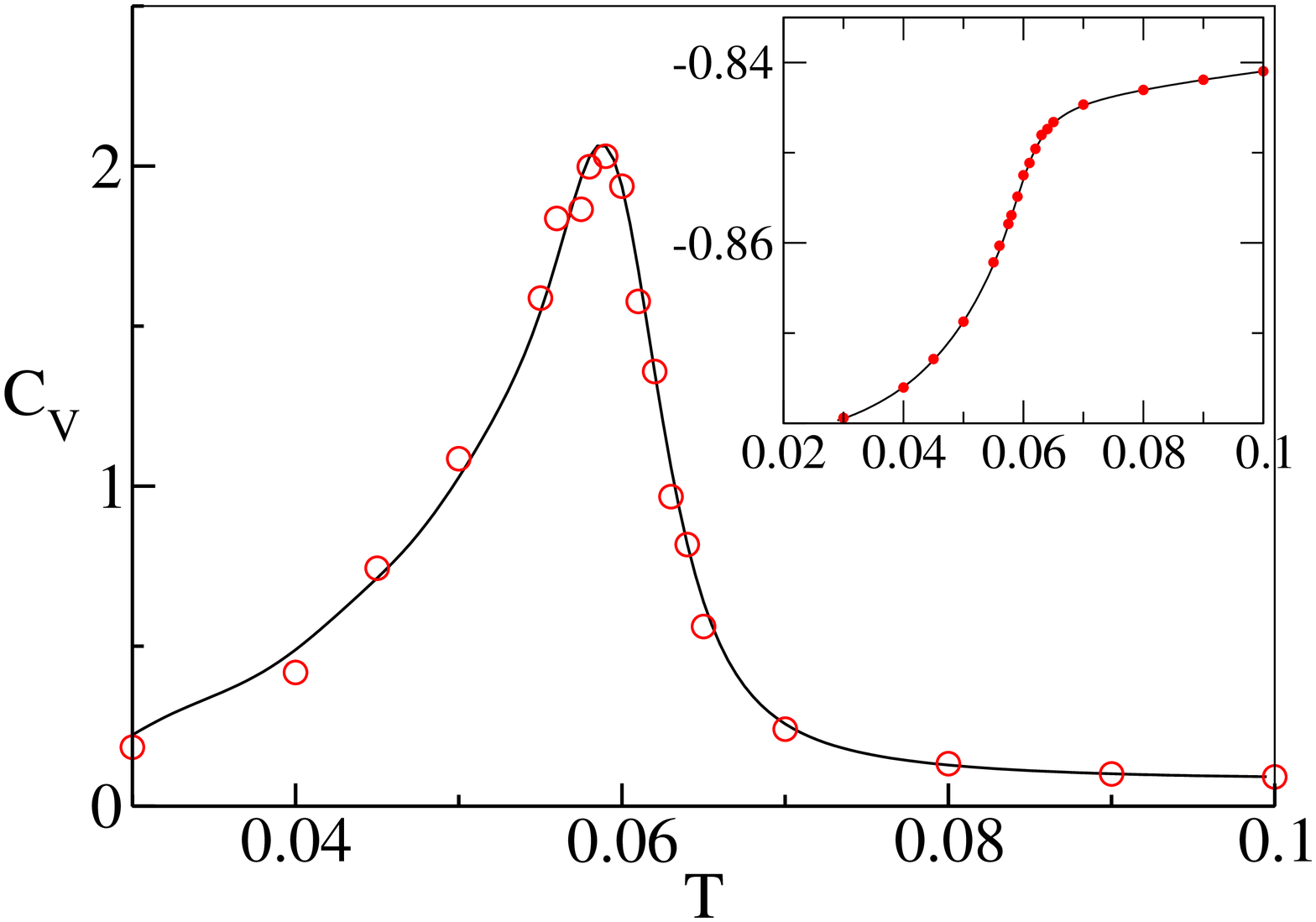}
  \end{minipage}
\begin{minipage}{.32\textwidth}
  \centering
   \includegraphics[width=1.1\columnwidth]{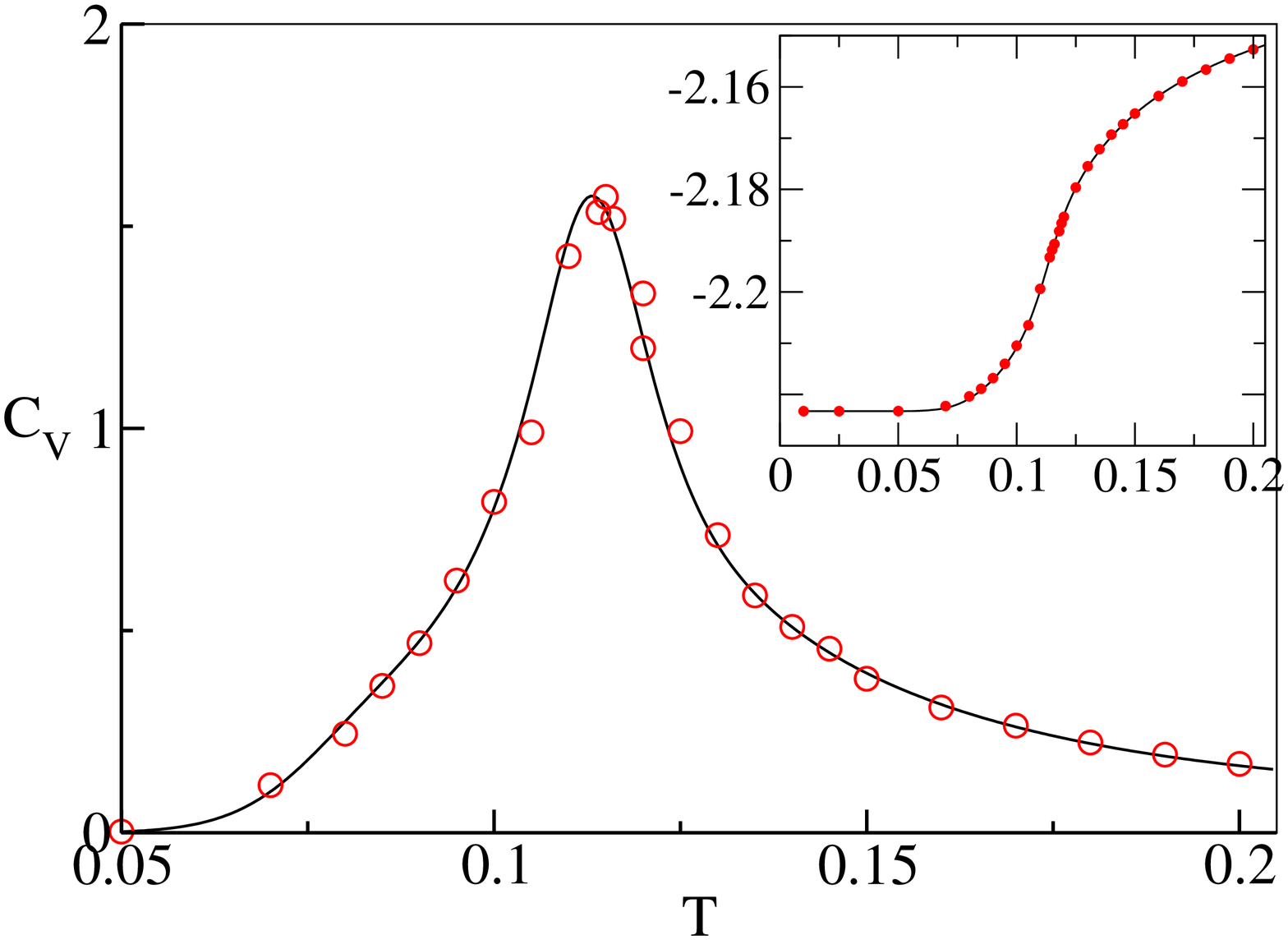}
 \end{minipage}
 \caption{(Color Online). Specific heat $C_V$ per site for the 2D Falicov-Kimball model at half-filling obtained with our (black line) and Metropolis (red circles) algorithms.  %The electron bandwith $W=4$ while the
  The electron-ion onsite repulsion $U/t = $ $0.25$ (weak-coupling), $1$ (moderate-coupling) and $8$ (strong-coupling) from the left to the right panel. Lattice size is 16x16 with PBC. Insets: comparison of internal energies.}
 \label{fig3}
\end{figure*}

We now highlight some important distinctions between the WL and our algorithms. 
A  single step in the second stage of our algorithm entails accumulation of full thermodynamic information from all configurations of subsystem $B$ (a remarkable $2^{1800}$ configurations in the presented simulation) during each update move on subsystem $A$. In contrast the standard WL algorithm updates information from one configuration per move. However this "exponential update" comes at the initial cost 
of first obtaining the "auxiliary" DOS via a WL procedure. Interestingly, the "auxiliary" DOS has to be determined with essentially a higher histogram flatness requirement than the system DOS in the direct WL simulation of the Ising model. This has two sources: (i)  The "subsystem Hamiltonian" with energies $\varepsilon_{\min}$  contains multi-spin and long range interactions unlike the original Ising interaction Hamiltonian. (ii) Sensitivity to the precision of the "auxiliary" DOS. We have found that adding small fluctuations to $\tilde{\rho}(\epsilon_{\rm min})$ rapidly deteriorates results. 
Hence, our algorithm, should not be viewed as an alternative to the standard WL algorithm for \textit{classical} models.
Finally while the WL algorithm  directly outputs the system DOS, from which the partition function may be easily calculated, our algorithm  yields the partition function. 
This  difference  has important practical consequences. In the WL algorithm the DOS values (which generally grow exponentially  with system size) are accumulated in a given run by {\it multiplying} them with the so-called modification factor $f$ every time a given energy is visited. In subsequent runs $f$ is gradually reduced to 1. This ingenious trick allows to build up very big DOS values in a reasonable number of steps). In our procedure we {\it sum} the accumulated quantities $\exp(-\beta F(\beta|A))/w(A)$, and the values of both factors are {\it already} very big or very small.  Therefore  care must be taken  to avoid roundoff error acumulation during summation. 

To illustrate that our algorithm succesfully simulates discontinuous phase transitions, we consider the 10-state Potts model on the square lattice with nearest neighbour interaction. Bi-separability here may be shown in  similar fashion as in the Ising model. However the form of the conditional partition function is more complicated than (\ref{ZIsing}) due to the multitude of values of local fields set by configurations of nearest neighbour Potts spins.  Comparison in Fig. \ref{fig2} of the specific heat, obtained by the WL and  our algorithm, for a 30 x 30 -- site lattice with PBC reveals the effectiveness of our sampling scheme in this case as well.

Now,  consider the Falicov-Kimball Hamiltonian (FKH)   of FCCF:
\begin{eqnarray}\label{FKH}
H_{FK}=-t\sum_{<i,j>}c^{\dagger}_i c_j + U\sum_{i} n^{c}_i n^{d}_i - \mu\sum_{i} n^{c}_i, 
\end{eqnarray}
where $ c^{\dagger}_i $ and $ d^{\dagger}_i $ are creation operators of mobile and immobile fermions respectively, $n^{c}_i= c^{\dagger}_i c_i $, $ n^{d}_i = d^{\dagger}_i d_i$, $t$ is the nearest neighbour hopping integral, $U$ is the Hubbard on-site interaction and $\mu$ is the chemical potential for $c$ fermions (the number $N_d$ of immobile fermions is fixed). For a given configuration of immobile fermions $\{ n^{d}_i \}$, the set of single particle eigenenergies $\{ \varepsilon_l \}$ of mobile fermions is readily obtained, rendering the model  bi-separable.

Unlike the classical models above, an efficient standard WL algorithm is not directly applicable to the simulation of the FKH.
We consider the FKH on a $N=L^2$ square lattice under PBC with $L=16$, at half filling ($\mu=U/2, N_d=N/2$), for different values of $U$.
Under these conditions, the FK model undergoes, at low temperatures, a transition to the charge density wave (CDW) ordered state, with $Q=(\pi,\pi)$ ordering wave-vector. The transition is continuous for large $U$ with some evidence pointing to a change to discontinous transitions for small $U$ \cite{Maska,Zonda}. The application of our algorithm yields all-temperature results unlike standard Metropolis sampling over the immobile fermions. Moreover, since the Metropolis algorithm suffers from slow kinetics at discontinous phase transitions our method should be particulary useful in further investigations of the small $U$ regime. Here, we restrict to proving that our sampling principle works. In Fig. \ref{fig3}, a comparison of  results for the specific heat and internal energy obtained by  Metropolis sampling  and  our algorithm is presented. On all diagrams the results are in good agreement within the accuracy of the local update Metropolis algorithm, which again indicates the effectiveness of our sampling. We mention here, that in obtaining the "auxiliary" DOS in the first stage of our algorithm, we discretized values of minimal energies of the total system (but we have not discretized single particle energies anywhere else) by assigning them to (energy) bins of width  $0.005t$ or $0.001t$ and checked convergence.

{\it Summary and conclusions - } 
In this Letter, we have presented a new Monte Carlo algorithm for problems of fermions coupled to classical degrees of freedom. This algorithm is based on Wang-Landau-like sampling in contradistinction to the commonly used Metropolis sampling. This allows in principle to overcome drawbacks of Metropolis schemes and  importantly to  fully exploit {\it all} available thermodynamic information at each diagonalization step --  information that is mostly wasted in other MC schemes. The scheme is based on the notion of {\it minimal energy attainable for a given classical field configuration}. As a minimal requirement, we devised a rule that all such minimal energies be visited by the algorithm. This was achieved by first obtaining  an auxiliary DOS for minimal energies via Wang-Landau sampling, and then  using its inverse as the weight function to perform a random walk in  classical field configurations  accumulating the full partition functions. We benchmarked our recipe for several paradigmatic models of statistical physics achieving excellent agreement with known results. We mention here that while our  principle of sampling minimal energies yields satisfactory results in the presented examples, more generally, supplementary conditions may need to be identified in other cases. However, our results show that in principle accumulation of conditional partition functions  at all temperatures at once using simple, temperature independent importance sampling is possible. An intriguing possibility is the application of this method to problems of quenched disorder, where free energies need to be efficiently averaged over the disorder realizations.

Finally, we comment on the most distinctive feature of our algorithm -- {\it i.e.} accumulation of full conditional partition functions per update.
Standard algorithms update information from only single configurations per move. However, this is by no means the only possibility.  N-fold way \cite{nfold} algorithms may be seen as updating information from $M$ configurations during each move (where $M$ is the number of system sites). Our algorithm is remarkable in that it uniquely allows the update of  information from an exponential number of configurations during each move ($\sim \exp (\alpha M)$ for some constant $\alpha$). Clearly,  no algorithms exist that can update more information from a complexity point of view.

\begin{acknowledgements}
{\it Acknowledgements - } We thank Mathias Troyer for interesting and useful comments. The authors acknowledge support from the (Polish) National Science Center Grant No DEC-2011/03/B/ST2/01903.
\end{acknowledgements}

\end{document}